\documentstyle[multicol,aps,epsf]{revtex}
\def\E{\varepsilon}
\def\rmd{{\rm d}}
\def\rme{{\rm e}}
\newlength{\ul}
\newcommand{\PA}{\!\!\!\raisebox{-0.4\ul}
{\setlength{\unitlength}{0.1\ul}
\begin{picture}(11,10)(0,0)
\thicklines
\put(1,5){\put(0,0){\circle*2}
\put(5,0){\circle{10}}
}\end{picture}}}
\newcommand{\PB}{\!\!\!\raisebox{-0.4\ul}
{\setlength{\unitlength}{0.1\ul}
\begin{picture}(26,10)(0,0)
\thicklines
\put(16,5){
\put(-15,0){\circle*2}
\put(-5,0){\dashbox{0.5}(5,0){}}
\put(-5,0){\circle*2}
\put(0,0){\circle*2}
\put(5,0){\circle{10}}
\put(-10,0){\circle{10}}
}\end{picture}}}
\newcommand{\PG}{\!\!\!\raisebox{-0.4\ul}
{\setlength{\unitlength}{0.1\ul}
\begin{picture}(16,10)(0,0)
\thicklines
\put(1,5){
	\put(0,0){\circle*2}
	\put(7.5,5){\circle*{2}}
	\put(7.5,-5){\circle*{2}}
	\put(7.5,0){\oval(15,10)}
	\put(7.5,-5){\dashbox{0.5}(0,10){}}
}\end{picture}}}

\newlength{\breite}
\begin{document}
\title{A geometric generalization of field theory  to manifolds of arbitrary dimension}
\author{Kay J\"org Wiese\\
\small Fachbereich Physik, Universit\"at GH Essen,  45117 Essen,
Germany}
\author{Mehran Kardar\\
\small Department of Physics, MIT, Cambridge, Massachusetts 02139, USA}
\maketitle
\begin{abstract}
We introduce a generalization of the $O(N)$ field theory to 
$N$-colored membranes of arbitrary inner dimension $D$. 
The $O(N)$ model is obtained for $D\to1$, while $N\to0$ leads to self-avoiding
tethered membranes (as the $O(N)$ model reduces to self-avoiding polymers).
The model is studied perturbatively by a 1-loop renormalization group analysis,
and exactly as $N\to\infty$.
Freedom to choose the expansion point $D$, leads to precise estimates
of critical exponents of the $O(N)$ model.
Insights gained from this generalization include a conjecture on the 
nature of droplets dominating the $3d$-Ising model at criticality;
and the fixed point governing the random bond Ising model. 

\medskip \noindent {PACS numbers: 05.70.Jk, 11.10.Gh, 64.60.Ak,  
75.10.Hk }
\end{abstract}
\begin{multicols}{2}\narrowtext
Field theories have strong connections to 
geometrical problems involving fluctuating lines.
 For example, summing over all world-lines representing
the motion of particles in space-time, is the Feynman
path integral approach to calculating transition probabilities, 
which can also be obtained from a quantum field theory.
Another example  is the high-temperature expansion of the Ising model,
where the energy-energy correlation function is a  sum over all 
self-avoiding closed loops which pass through two given points.
Generalizing from the Ising model to $N$  component spins,
the partition function of a corresponding $O(N)$ ``loop model'' is 
obtained by summing over all configurations of a gas of closed loops, 
where each loop comes in $N$ colors, or has a fugacity of $N$.
In the limit $N \to 0$, only a single loop contributes, giving the partition 
function of a closed self-avoiding polymer\cite{PGG72}.
  
There are several approaches to generalizing 
fluctuating lines to entities of other internal dimensions $D$.
The most prominent examples are string theories and lattice gauge
theories, which both describe $D=2$ world sheets\cite{LHLXII}.
The low temperature expansion of the Ising model in  $d$ dimensions also 
results in a sum over surfaces that are $d-1$ dimensional.
Each of these generalizations has its own strengths, and offers
new insights on field theory. Here we introduce a generalization
based on a class of $D$-dimensional manifolds called ``tethered" 
(or polymerized) membranes, which have fixed internal connectivity,
and are the simplest generalization of linear polymers\cite{KKN}.

{\em Tethered manifolds} are a quite successful generalization
of self-avoiding polymers embedded in $d$-dimensional external space:
Simple power counting indicates that self-avoidance is relevant only for 
dimensions $d<d_c=4D/(2-D)$, making possible an 
$\E=2D-d(2-D)/2\sim (d_c(D)-d)$-expansion, which was
first carried to 1-loop order  in Refs.~\cite{1loop}.
 Following more rigorous analysis of this novel perturbation series
\cite{sum}, recently 2-loop calculations were performed for 
$1\leq D<2$ \cite{2loop}.
To obtain results for polymers or membranes, one has the freedom
to expand about {\em any} inner dimension $D$, and the corresponding
upper critical dimension of
the embedding space\cite{Hwa90}.
This 
\begin{figure}[h]\setlength{\unitlength}{0.7mm}%
\centerline{\begin{picture}(80,50)(0,-10)
\thinlines
\put(0,30){\makebox(0,0){\begin{minipage}{5cm}\begin{center}
\small ${O}(N)$ field theory\end{center}\end{minipage}}}
\put(00,8){\makebox(0,0){\begin{minipage}{5cm}\begin{center}\small self-avoiding polymers
\end{center}\end{minipage}}}
\put(75,8){\makebox(0,0){\begin{minipage}{5cm}\begin{center}\small self-avoiding $D$-dimensional tethered membranes\end{center}\end{minipage}}}
\put(75,30){\makebox(0,0){\begin{minipage}{5cm}\begin{center}
\small ${M}(N,D)$ manifold model\end{center}\end{minipage}}}
\put(25,8){\line(1,0){20}}
\put(25,8.04){\makebox(0,0){\,\,\,$<$}}
\put(30,13){\makebox(10,0){\small $D\to 1$}}
\put(25,30){\line(1,0){20}}
\put(25,30.04){\makebox(0,0){\,\,\,$<$}}
\put(30,35){\makebox(10,0){\small $D\to 1$}}
\put(0,14){\line(0,1){10}}
\put(0.05,14.8){\makebox(0,0){$\vee$}}
\put(0.05,14.3){\makebox(20,10){\small $N\to 0$}}
\put(75,15){\line(0,1){10}}
\put(75.05,15.8){\makebox(0,0){$\vee$}}
\put(75.05,15.3){\makebox(20,10){\small $N\to 0$}}
\end{picture}}
\vspace{-4mm}%
\caption{Schematic depiction of the model, and its limits.}%
\label{scheme}%
\end{figure}
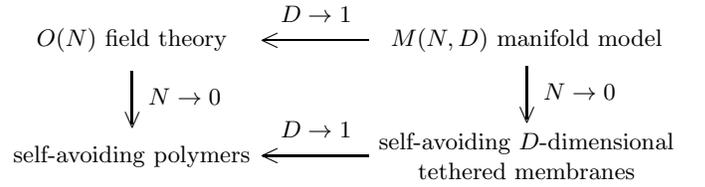%
\noindent
freedom can be used to optimize calculation of critical exponents,
and when  applied to polymers, the results are clearly better than those
from the standard expansion.

Our basic idea is to generalize the {\em high temperature expansion} 
of the $O(N)$ model, from a gas of self-avoiding loops of fugacity $N$, 
to a similar gas of closed fluctuating manifolds of internal dimension $D$.
The primary goal is to obtain a novel analytical handle on the field theory 
for $D=1$, and we do not insist that the models for general
$D$ correspond to any physical problem.
Given this caveat, the generalization is not unique.
Encouraged  by its success in polymer theory, we study {\em tethered} 
manifolds, and in addition restrict ourselves to the genus of  hyper-spheres.  
We have chosen hyperspheres as they have no additional anomalous correction 
{\em  exactly at } $D=2$. 
The resulting manifold theory depends on two parameters $N$ and $D$, whose
limiting behaviors reduce to well known models, as indicated in Fig.~\ref{scheme}.

We construct the $M(N,D)$ manifold model {\em perturbatively},
starting with the well known  high-temperature expansion (loop-model)
of the $O(N)$ field theory as
\begin{eqnarray} \label{pert}
\ln {\cal Z}&=& \,\PA \,+ \,\PB\,  + \,\PG\, + \cdots \ .
\end{eqnarray}
The first term in this sum is  the partition function of a non-interacting
polymer loop, now generalized to a $D$-dimensional
manifold, fixed at the origin in $d$ dimensions
\begin{equation}
\PA \,= \frac{c(D)}{D} \int_0^{\infty} \rmd \Omega \,\,
\Omega^{-\frac{2-D}{D}d} \rme^{-t \Omega}\ .
\end{equation}
The integral runs over all sizes $\Omega$, weighted by a 
chemical potential $t$, while the factor 
\begin{equation} 
\Omega^{-{\frac{2-D}D d }} \sim \left< \tilde \delta^{d}(r(x_0)) \right>_0 \ ,
\end{equation}
is the probability that the membrane is attached to a given point in space\cite{sum}.
The above generalization depends on a function $c(D)$, 
related to the relative strengths of 
self-avoidance between parts of the same manifold, 
and between different manifolds. Any choice of $c(D)$ which
reproduces the  polymer partition function with $c(1)=1$, is acceptable.
In the remainder, we will  mainly focus on $c(D)=D$, equivalent to the 
integral $\rmd \Omega$ over all scales. 
 
Subsequent terms in Eq.~(\ref{pert}) correct for the intersections 
(symbolized by a dashed line) between manifolds.
Configurations involving intersections between different manifolds, or
self-intersections of the same manifold (e.g., the second and third terms
in the above expansion, respectively)  have to be
subtracted from the partition function.
The contributions from these terms result in divergencies 
which have to be removed by renormalization
(for  details see Ref.~\cite{WieseKardar98a}). 
The final result is the 1-loop expansion of the exponent  $\nu(D,N,d)$
(for the divergence of the correlation length at a critical point), 
given by ($\E\equiv 2D-\frac{2-D}2d$)
\begin{eqnarray}\label{central}
&&\nu(D,N,d) ={2-D\over 2}\times\\
&& \left[1+{ \E\over 2D}
{1+{c(D)}{N\over2} \over {1\over 2-D} \Gamma\left(\frac D{2-D}\right)^{2}
{\Gamma\left(\frac{2D}{2-D}\right)^{-1}}+1 + c(D) {N\over4}}\right]\ .\nonumber
\end{eqnarray}
This expression reduces to the well-known $\E$-expansion\cite{Zinn}
around $d=4$ for lines ($D=1$), while the $N\to0$ limit reproduces the result for 
self-avoiding manifolds\cite{1loop}.

The perturbative series can also be summed exactly in the  {\em large $N$} limit,
where the ambiguity associated with $c(D)$ is removed\cite{WieseKardar98a}, to give
the exponent
\begin{equation}
\nu_{N\to\infty} = \frac{D}{d-{2D}/{(2-D)}}\ .
\end{equation}
This generalizes the  well-known result of $\nu=1/(d-2)$ for the $O(N)$ model\cite{Zinn}.

To demonstrate the utility of this generalization, we next estimate the 
exponent $\nu$ in physical dimensions, 
by  expanding about a point  $(D_0,d_0)$ 
on the critical curve  $\E(D_0,d_0)=0$.
The simplest scheme is to extrapolate towards the physical theories
for $D=1,2$ and $d=2,3,\cdots$, with the
expansion parameters
$D-D_0$ and $d-d_0$. 
However,  this choice is not optimal (see Ref.~\cite{2loop}), 
and better results are obtained with expansion parameters
$D_c(d)={2d}/{(4+d)}$ and $\E(D,d)=2D -d {(2-D)}/2 d$. 
Furthermore, it is better to expand 
$\nu d$ or $\nu (d+2)$ rather than  $\nu$.
The first is an expansion of the form $\nu d =2 D + a(D) \E $,
about the {\em mean-field} result $\nu_{\mbox{\scriptsize MF}}= {2D}/{d} $, 
also  known as Gaussian variational approximation\cite{Variational}.
The second is a similar expansion about the {\em Flory} expression
$\nu_{\mbox{\scriptsize Flory}} = (2+D)/(2+d)$.
\begin{figure}[h]
\centerline{\epsfxsize=\breite \parbox{\breite}{\epsfbox{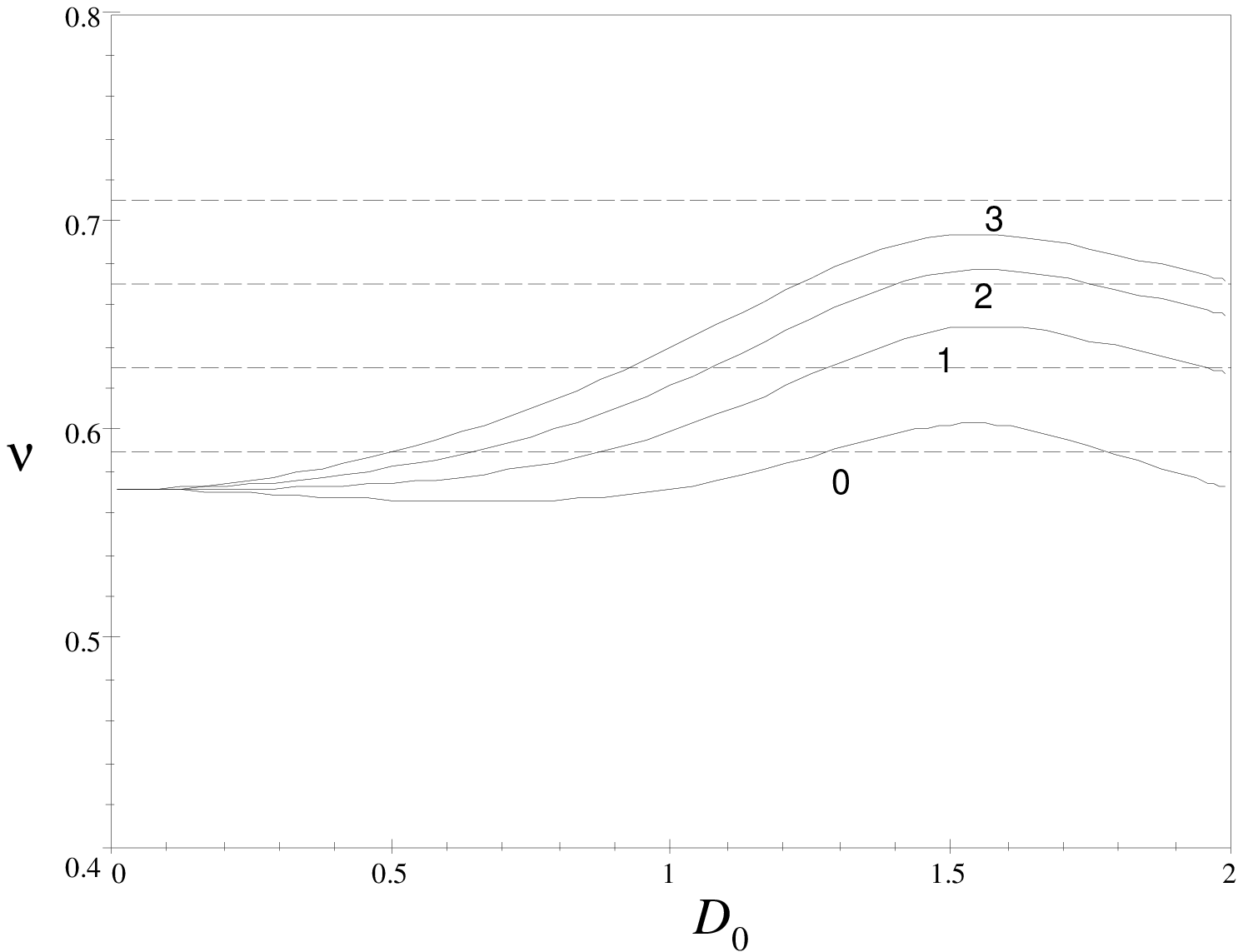}}}\vspace{-2.6cm}%
\centerline{\tabcolsep1.4mm
\renewcommand{\arraystretch}{1.20}
\hspace{14mm}\begin{tabular}{|c|c|c|c|c|} \hline
 $N$ & 0 &1 &2&3 \\ \hline 
 $\nu$, our result & 0.601 & 0.646 & 0.676 & 0.697 \\ \hline
 $\nu$, from \cite{Zinn}  & 0.589 & 0.631 & 0.676 & 0.713 \\ \hline
\end{tabular}\renewcommand{\arraystretch}{1.0}}\vspace{11mm}
\caption{Extrapolations for the exponent $\nu$ of the 
$O(N)$ model in $d=3$, using the expansion of $\nu d$ with $c(D)=D$.
The dashed lines  represent the best known
values from Ref.~\protect\cite{Zinn}.
}\label{extr}%
\end{figure}%

After selecting one of these schemes, Eq.~(\ref{central})  is re-expressed
in terms of the chosen expansion parameters.
However, we are still free to choose the expansion point along the 
critical curve, which then fixes $D_0$. 
As $D_0$ is varied, different values for $\nu$
are obtained, as ploted in Fig.~\ref{extr}.
The criterion for selecting a particular value from such curves 
is that of minimal sensitivity to the expansion point, and we thus 
evaluate $\nu$ at the extrema.
The broadness of the extremum then provides a measure for the 
goodness of the result, and the expansion scheme. 
Although we examined several such curves, only a selection is reproduced in 
 Fig.~\ref{extr}. 
Our results are clearly better than the standard 1-loop expansion
of $ \nu =1/2 +{(N+2)}/[{4 (N+8)]}$.

While providing good estimates for exponents of the $O(N)$ model
is certainly a benefit, the generalization to $M(N,D)$ should offer
insights beyond the standard field theory. 
Also, for the approach to be more generally applicable, we should show 
that similar generalizations are possible for other field theories. 
In the rest of this article we shall demonstrate that these
goals are indeed feasible. 

Our starting point was the high temperature expansion of  the $O(N)$ spin model,
which naturally leads to a sum over $N$-colored loops ($D=1$); 
motivating the later generalization to $M(N,D)$.
 For the Ising model ($N=1$), a different geometrical description 
is obtained from a low temperature expansion: 
Excitations to the uniform (up or down pointing) ground
state are droplets of spins of opposite sign. 
The energy cost of each droplet is proportional to its boundary, 
i.e.\ again weighted by a Boltzmann factor $\rme^{-t\Omega}$.
Thus, a  low temperature series for the $d$-dimensional Ising partition function
is obtained by  summing over closed surfaces of dimension $D=d-1$. 
 For $d=2$, the high and low temperature series are similar, due to 
self-duality. For $d=3$, the low temperature description
is a sum over surfaces.
\begin{figure}[htb]\centerline{%
\epsfxsize=0.5\breite \parbox{0.5\breite}{\epsfbox{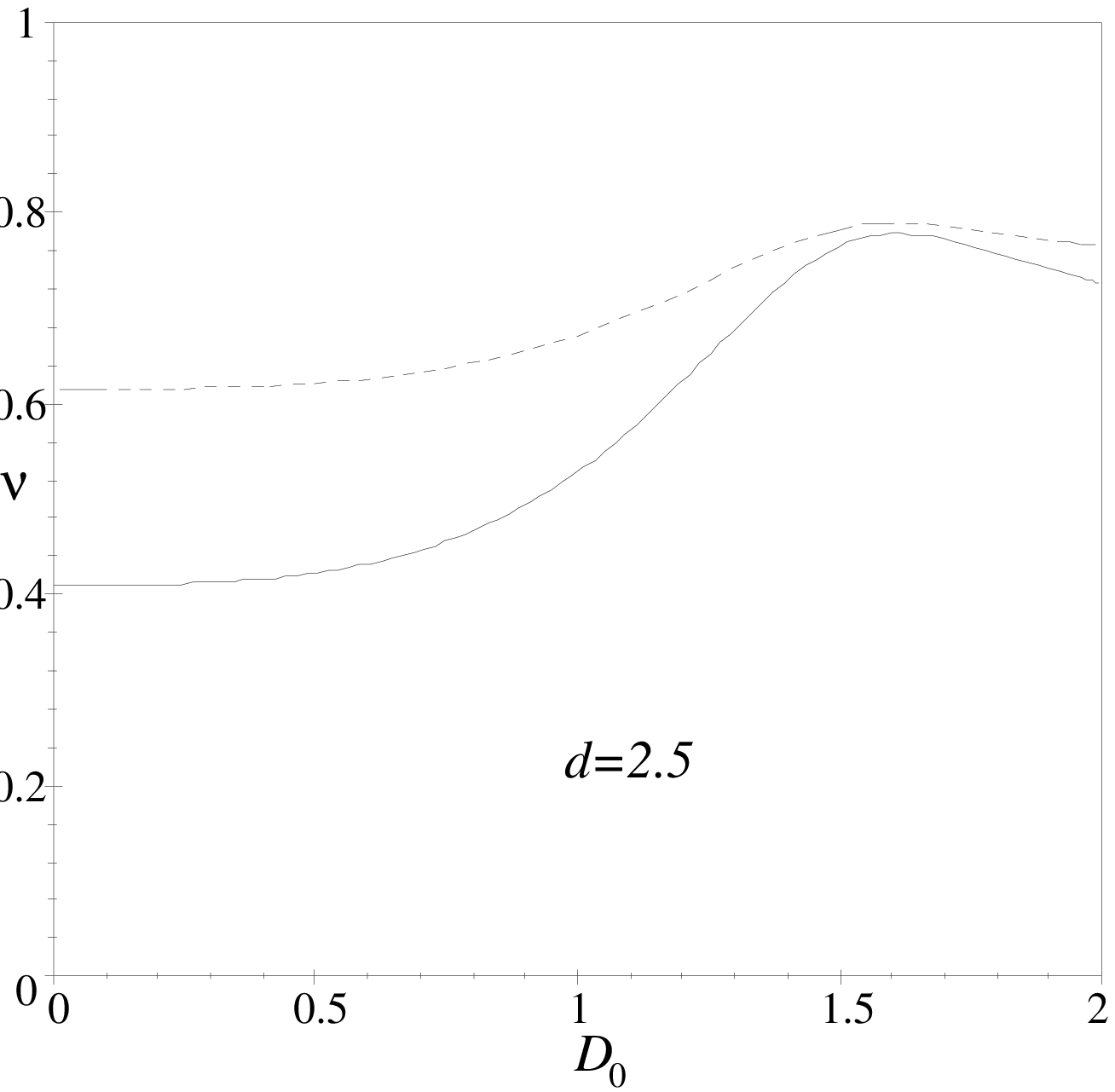}}
\epsfxsize=0.5\breite \parbox{0.5\breite}{\epsfbox{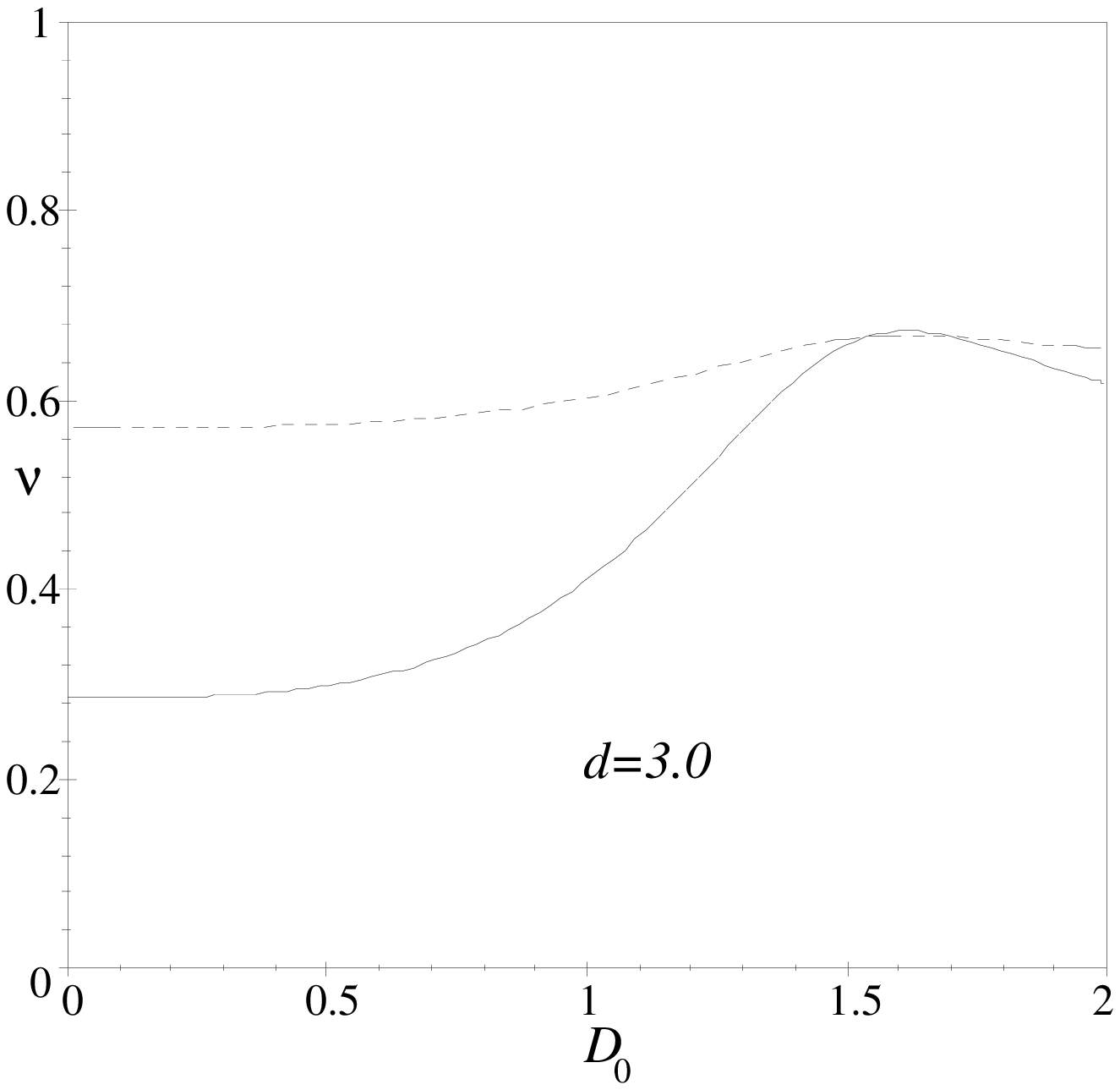}}}%
\caption{
Test of  Eq.~(\protect\ref{nu hypoth1}) for Ising models
in $d=2.5$ and $3$. The upper curves are from the
high temperature representation  ($D=1$), 
while the lower curves are from the low temperature description ($D=d-1$).  
The exponent $\nu$ is estimated from the maximum of each curve,
which are obtained by extrapolating $\nu d$
with $c(D)=D$, linearizing in $N$,  and dividing the result by $dD$.}
\label{specIsing}
\end{figure}%

The non-trivial question is regarding the types of surfaces which 
dominate the above sum. 
Since there is no constraint on the internal metric, 
 it may be appropriate to examine  {\em fluid membranes}. 
However, there is currently no practical scheme for treating interacting 
fluid  membranes, and the excluded volume interactions  are 
certainly essential to the problem. 
It is known that configurations of a single surface for $N=0$,
self-avoiding or not, are dominated by tubular shapes (spikes) which 
have very large entropy\cite{Cates}.
Such ``branched polymer'' configurations are very 
different from tethered surfaces.
However, for $N\ne 0$, it may be entropically advantageous to 
break up a singular spike into a string of many bubbles.
If so, describing the collection of bubbles by fluctuating 
hyper-spherical (tethered) manifolds may not be too off the mark\cite{topology}.
We can test the validity of this conjecture by comparing
the predictions of the dual high and low-temperature descriptions.

Singularities of the partition function are characterized by the critical
exponent $\alpha(D,d,N)$, or (using hyperscaling)  through
$
	\ln {\cal Z}_{\mbox{\scriptsize singular}} \sim |t-t_c|^{\nu d/D}.
$
The equality of the singularities on approaching the critical point from 
low or high temperature sides, leads to a putative identity
\begin{equation}
 \label{nu hypoth1}
{\nu(1,d,1)} =  \frac{1}{d-1}\,\nu(d-1,d, 1)
\ .
\end{equation}
Numerical tests of the conjecture in Eq.~(\ref{nu hypoth1}) 
are presented in Fig.~\ref{specIsing}. 
The extrapolated exponents (the maxima of
the curves) from the dual 
expansions are in excellent agreement. Nevertheless, higher-loop calculations would be useful to check
this surprising hypothesis.%

The simplest extension of the $O(N)$ model breaks the rotational
symmetry by inclusion of cubic anisotropy\cite{Amit}.
In the field theory language, cubic anisotropy is represented by a
term  $u\sum_i s_i^4$, in addition to the usual interaction
of $b\sum_{ij}s_i^2s_j^2$.
In the geometric prescription of 
high temperature expansions, the
anisotropic  coupling $u$ acts only between membranes 
of the same color, while the interaction $b$ acts  irrespective of color.
\begin{figure}[t]
\centerline{ \epsfxsize=\breite \parbox{\breite}{\epsfbox{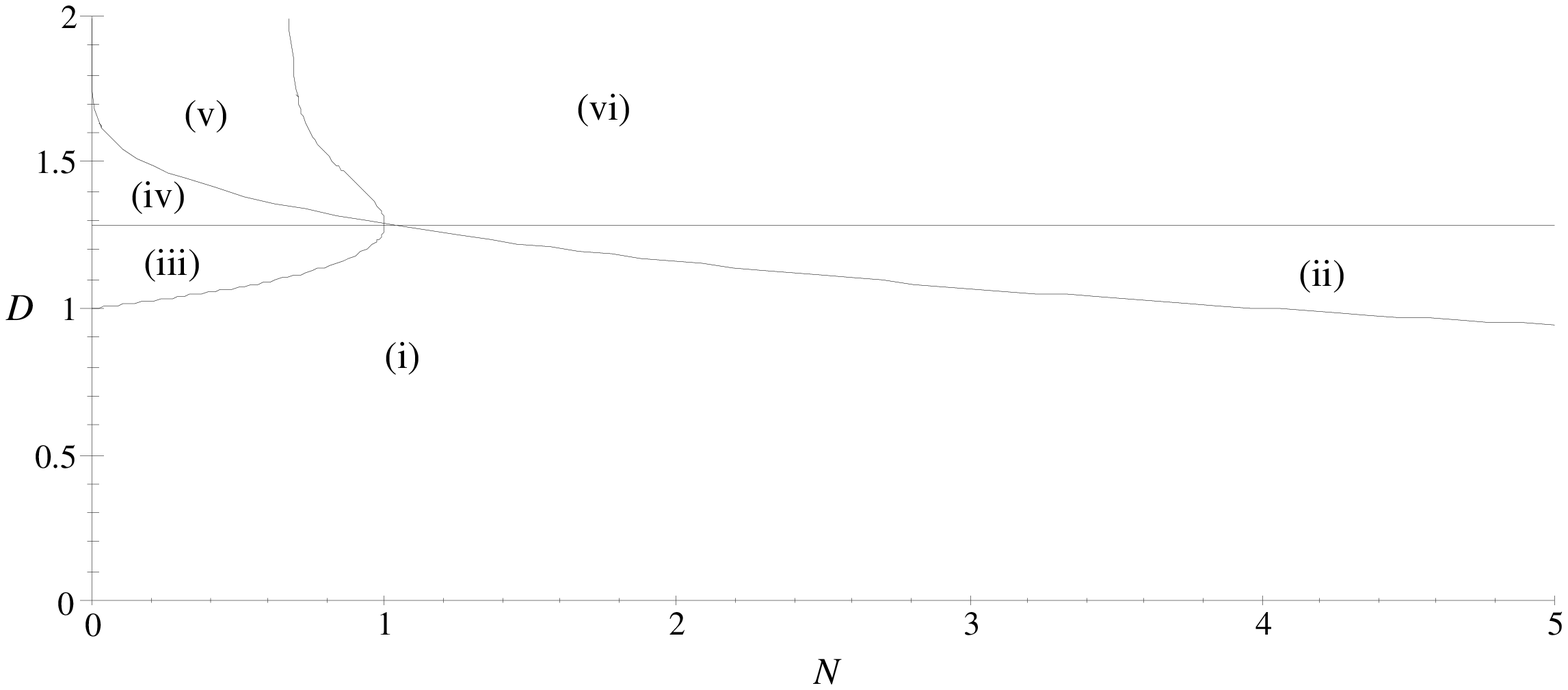}} } 
\centerline{%
\epsfxsize=0.33\breite \parbox{0.33\breite}{\epsfbox{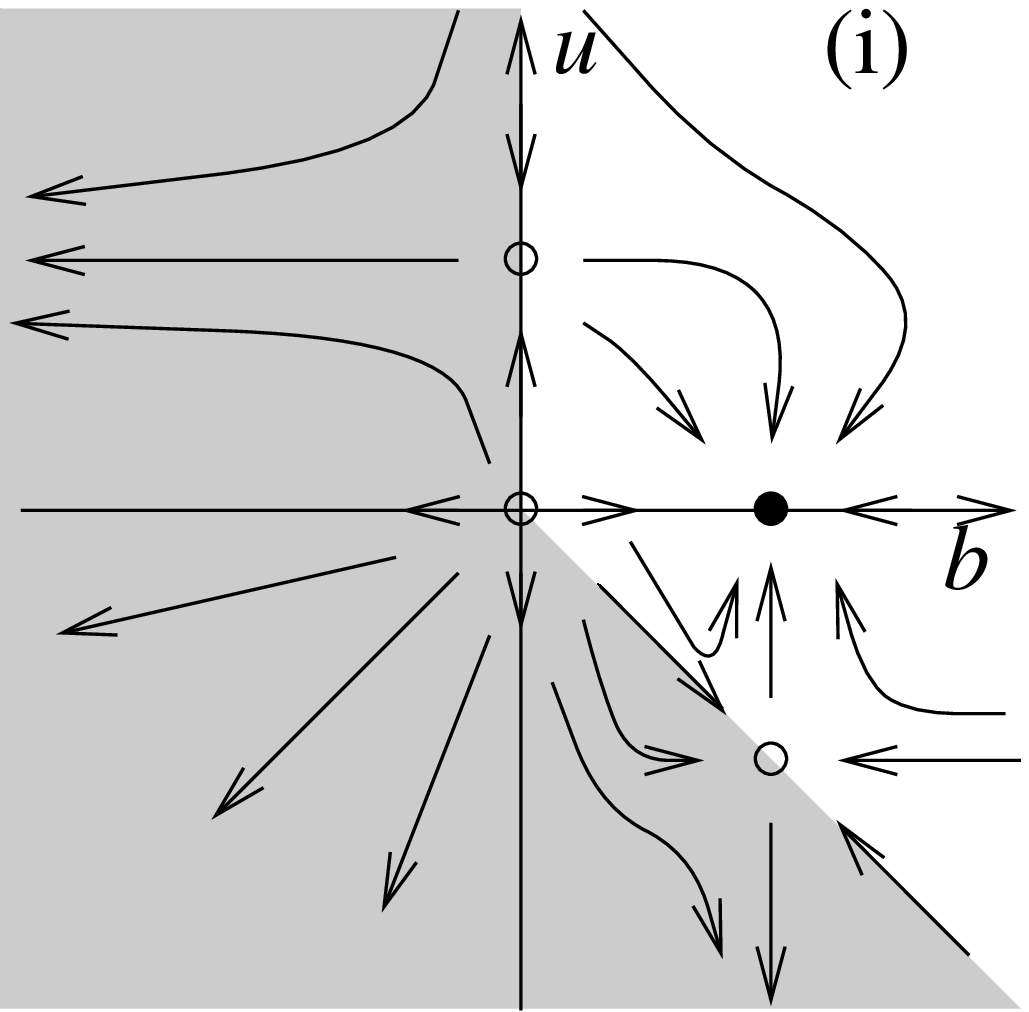}}%
\epsfxsize=0.33\breite \parbox{0.33\breite}{\epsfbox{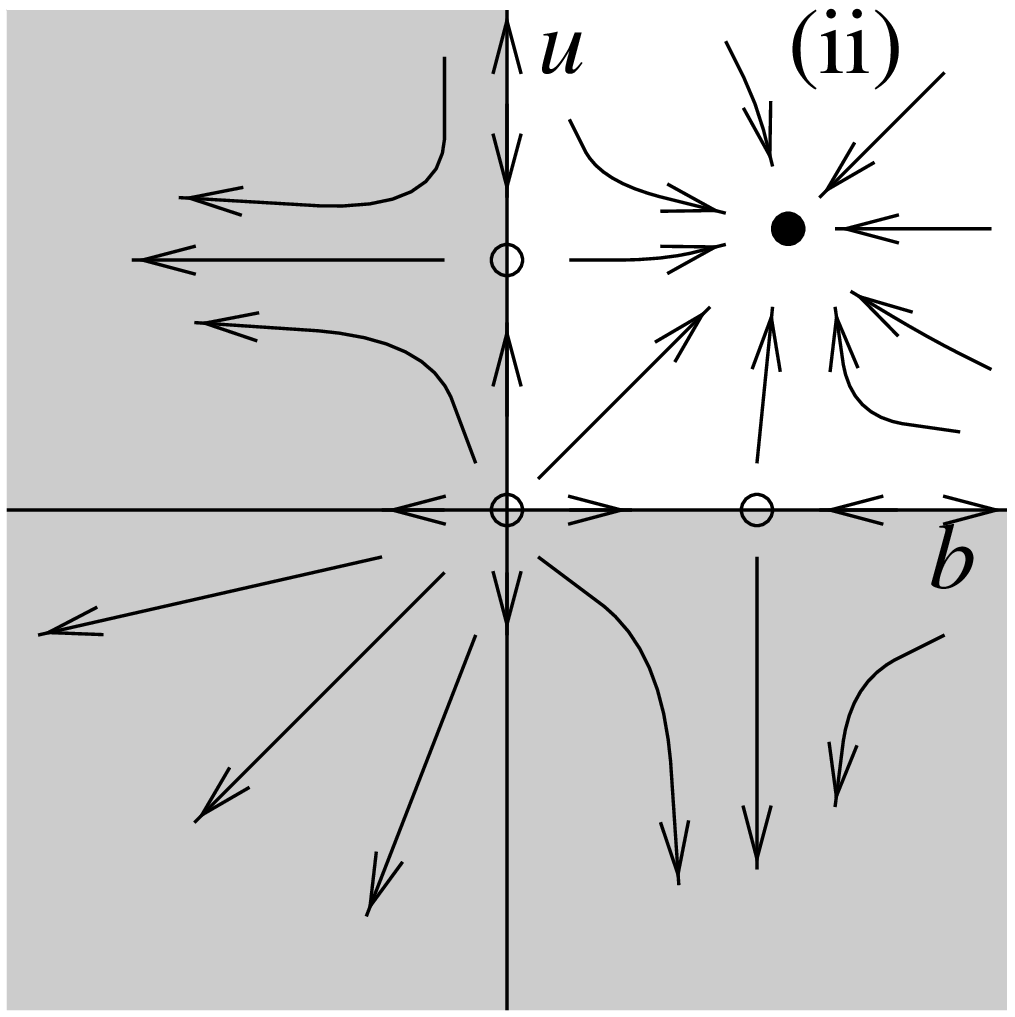}}%
\epsfxsize=0.33\breite \parbox{0.33\breite}{\epsfbox{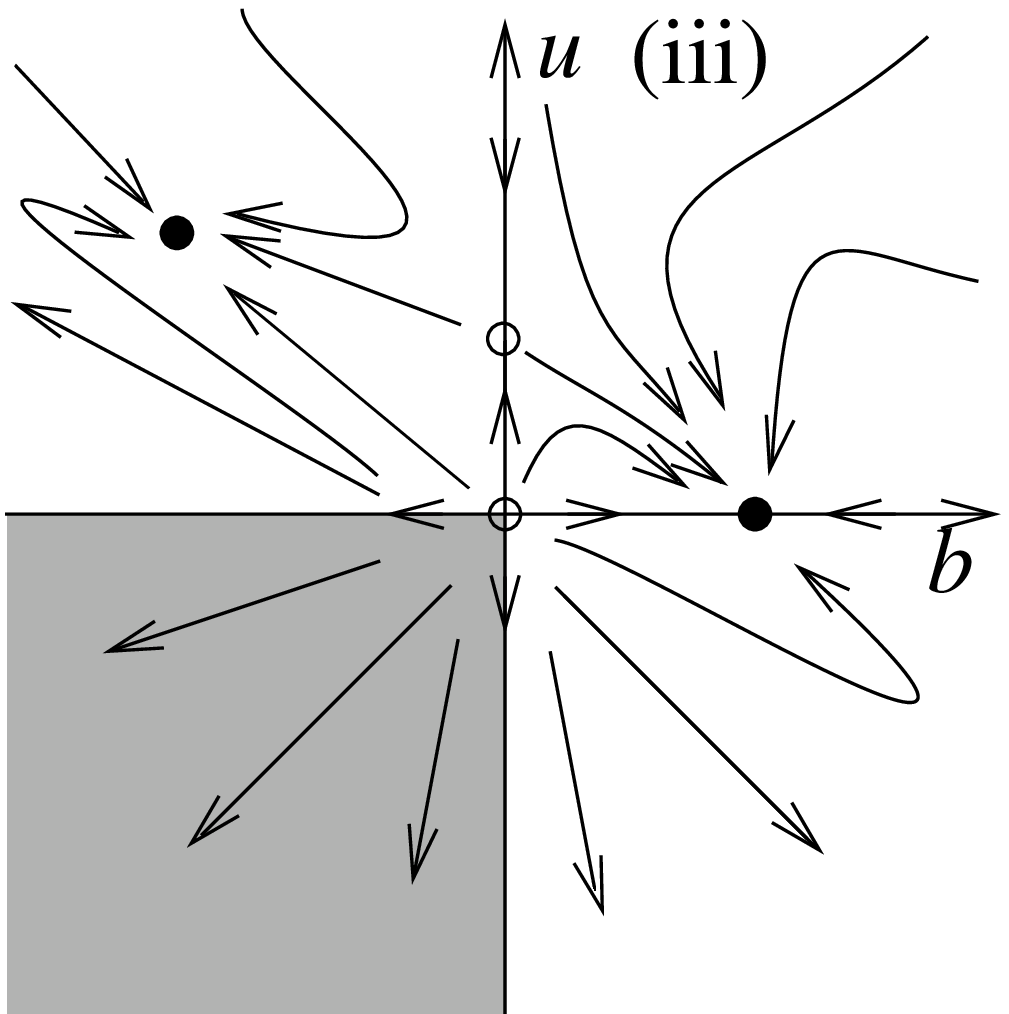}}}%
\centerline{%
\epsfxsize=0.33\breite \parbox{0.33\breite}{\epsfbox{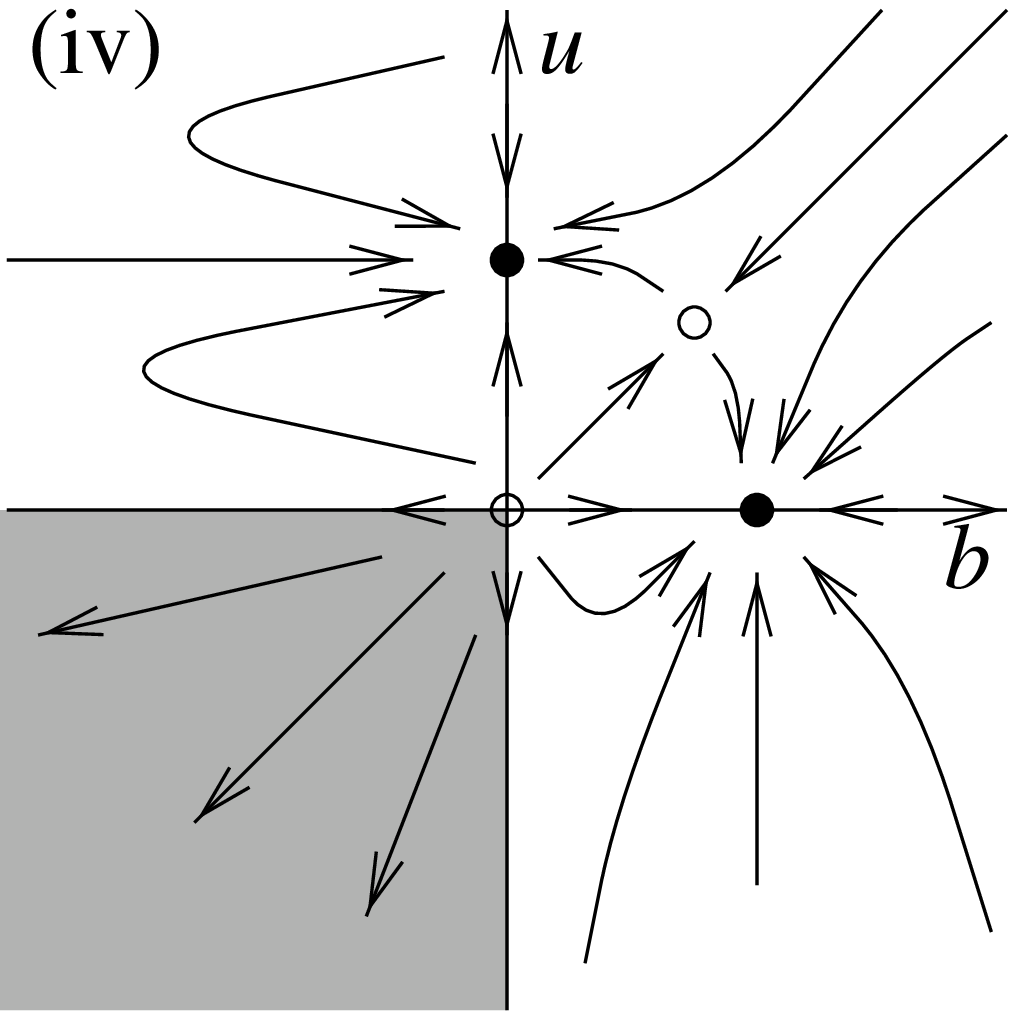}}%
\epsfxsize=0.33\breite \parbox{0.33\breite}{\epsfbox{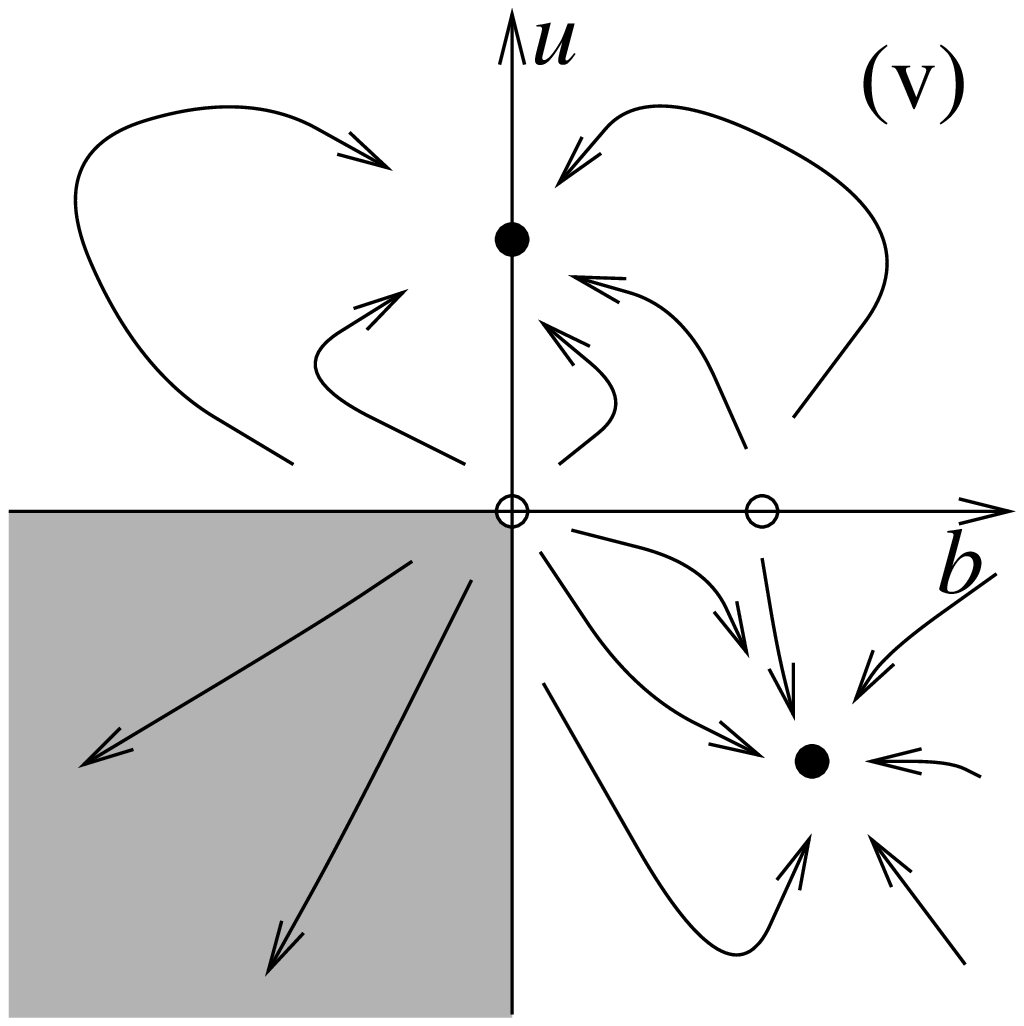}}%
\epsfxsize=0.33\breite \parbox{0.33\breite}{\epsfbox{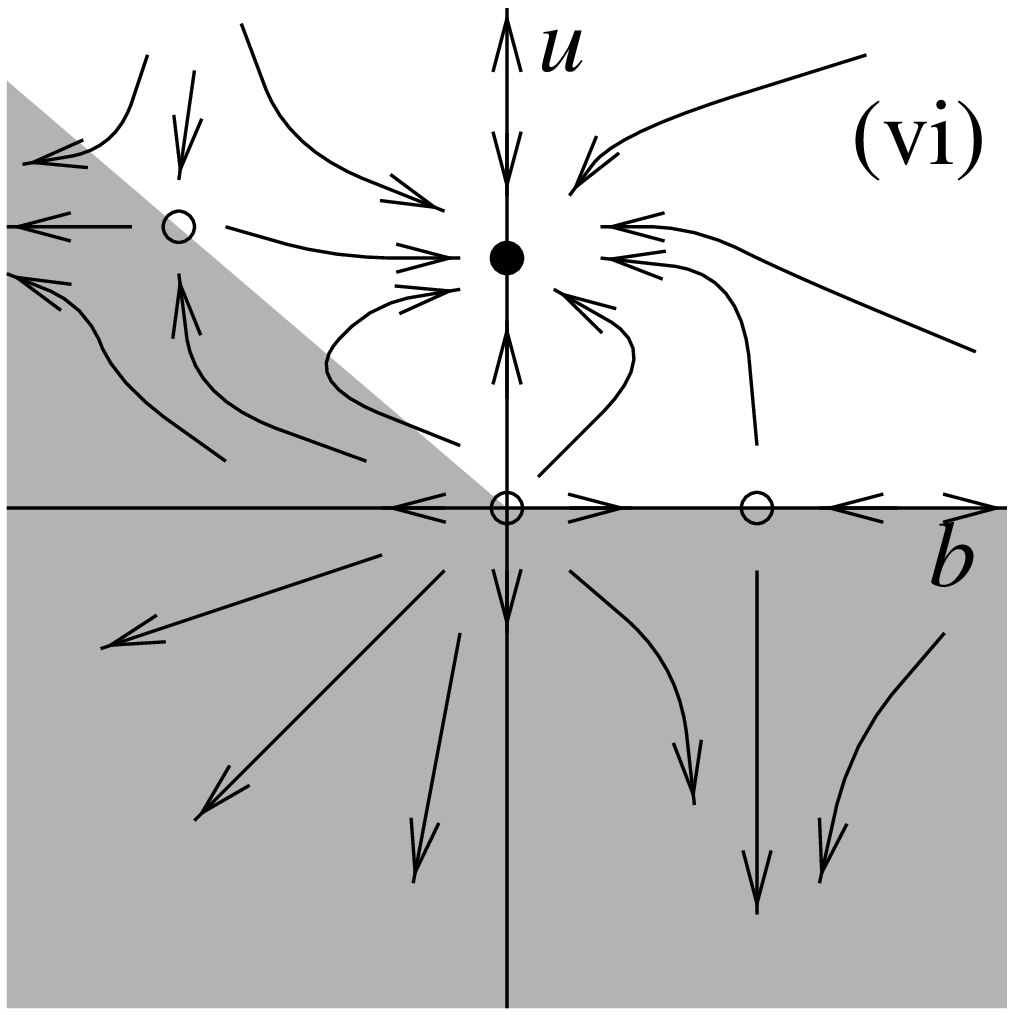}}%
}%
\caption{Regions with different RG-flow patterns in the $(N,D)$-plane (top),
and the corresponding RG-flows (bottom);
shaded regions are unstable.}\label{domains+flows}%
\end{figure}%

Stability of the system of colored membranes places constraints
on possible values of $b$ and $u$.
To avoid collapse of the system, energetic considerations imply that
if  $u<0$, the condition $u+b>0$ must hold, 
while if $u>0$, we must have $ u+Nb>0$\cite{Amit,WieseKardar98a}.
The above stability arguments are  expected to be modified upon the 
inclusion of fluctuations. A well-known example is the
Coleman-Weinberg mechanism \cite{Amit},
where the RG flows take an apparently stable combination of $b$ and
$u$ into an unstable regime, indicating that fluctuations 
destabilize the system.
In the flow diagrams described below, we also find the reverse behavior
in which an apparently unstable combination of $b$ and $u$ flows to
a stable fixed point. We interpret this as indicating that fluctuations
stabilize the model, a reverse Coleman-Weinberg effect,
which to our knowledge is new.
We have shaded in grey, the unphysical regions in the  flow diagrams 
of Fig.~\ref{domains+flows}.
As in their $O(N)$ counterpart, the RG equations admit 4 fixed points:
The {\em Gaussian} fixed point with $b^*_G=u^*_G=0$; 
the {\em Heisenberg} fixed point located at $b^*_H\neq0$, $u^*_H=0$; 
the {\em Ising} fixed point with $b^*_I=0$, $u^*_I\neq0$;
and the {\em cubic} fixed point  at $b^*_c\neq0$, $u^*_c\neq0$.
Furthermore, as depicted in Fig.~\ref{domains+flows}, there are 
six different possible flow patterns.
In the $O(N)$ model, the flows in (i) and (ii) occur for
$N<4$ and $N>4$, respectively.
The other four patterns do not appear in the standard field theory, 
as is apparent from
their domain of applicability in the $(N,D)$ plane
on Fig.~\ref{domains+flows}.
Note that there are  {\em two stable fixed points}
in three out of these four cases. 
\begin{figure}[t]
\centerline{ \epsfxsize=\breite \parbox{=\breite}{\epsfbox{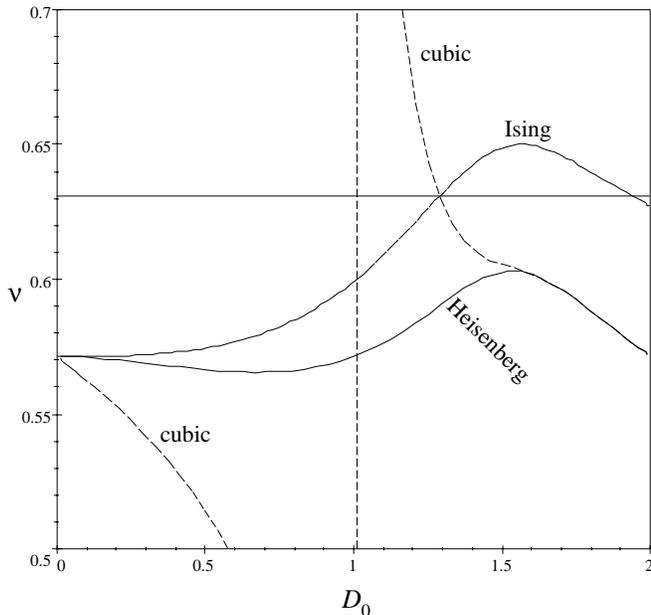}}}
\caption{Extrapolations of $\nu$ from the expansion of
$\nu d$ with $c(D)=D$, for the $O(N)$ model in $d=3$.
Exponents at the Heisenberg fixed point for $N=0$ are compared 
to those of the Ising and cubic fixed points. 
The crossing of the latter curves yields an estimate of 
$ \nu = 0.6315 $ for the $3d$ Ising model.
}
\label{extra KIH}
\end{figure}%

The $N\to0$ limit of the above models is interesting,  
not only because of its relevance to self-avoiding polymers and membranes, 
but also for its relation to the Ising model with bond disorder. 
The latter connection can be shown by starting with the field theory 
description of the random bond Ising model, replicating it $N$ times,
and averaging over disorder~\cite{RBIM}.
The replicated system is  controlled by a  Hamiltonian with
positive cubic anisotropy $u$, but negative $b=-\sigma$ ($\sigma$
is related to the variance of bond disorder).
From the  ``Harris criterion"\cite{Harris74}, new critical behavior
is expected for the random bond Ising system.
But in the usual field theory treatments\cite{RBIM},
there is no fixed point at  the 1-loop order.
In our generalized model, this is just the borderline between cases (i) and (iii).
However, we now have the option of searching for a stable fixed point 
by expanding about any $D\ne1$.
Indeed, for  $N=0$ and $1<D<1.29$, the cubic fixed point lies in the 
upper left sector ($u>0$ and $b<0$) and is completely stable,
as in flow pattern (iii). 

The extrapolation for $\nu$ at the cubic fixed point is plotted 
in Fig.~\ref{extra KIH}, where it is compared to the 
results for the Heisenberg and Ising fixed points. The divergence
of $\nu$ on approaching $D=1$ from above, is due to 
the cubic fixed point going to infinity as mentioned earlier. 
Upon increasing $D$, the Ising and cubic fixed points approach,
and merge for $D=1.29$.
 For larger values of $D$, the cubic fixed point is to the right 
of the Ising one ($b_c^*>0$), and only the latter is stable. 
Given this structure, there is no plateau for a numerical
estimate of the random bond exponent $\nu_{\mbox{\scriptsize DO}}$, 
and we can only posit the inequality 
$\nu_{\mbox{\scriptsize DO}} > \nu _{\mbox{\scriptsize Ising}}$.
While this is derived at 1-loop order, it should also hold at higher orders
since it merely depends on the general structure of the RG flows.
One may compare this to four loop calculations of the random bond 
Ising model \cite{Mayer89}, which are consistent 
with $\alpha=0$, i.e. at the border-line of the 
Harris criterion\cite{Harris74}, with $\nu=2/3$.

It is a pleasure to thank F.\ David, H.W.\ Diehl, and  
L.\ Sch\"afer for useful discussions.  Part of this 
work was done during a visit of K.J.W.\ to MIT, with
the   financial support  of  Deutsche Forschungsgemeinschaft,
through the Leibniz-Programm. 
The work at MIT is supported by the NSF Grant No.~DMR-93-03667.

\end{multicols}


\begin{references}
\bibitem{topology}
Another issue is whether the sum may be restricted to spheres, or if
 objects of other topologies must also be included. 
We argue in Ref.~\cite{WieseKardar98a} that the dominant contribution 
(and the only one included in perturbation theory) comes from spheres.
\end{references}
\end{document}